\documentstyle[12pt,aaspp4]{article}

\slugcomment{To be published in {\em The Astrophysical Journal},
             Feb. 10 1996}

\lefthead{MUSLIMOV \& PAGE}
\righthead{EVOLUTION OF VERY YOUNG PULSARS}

\newcommand{\be}{\begin{equation}}
\newcommand{\ee}{\end{equation}}
\newcommand{\Msol}{\mbox{$M_{\odot}\;$}}
\def\lsim{\lower 2pt \hbox{$\, \buildrel {\scriptstyle <}\over
         {\scriptstyle \sim}\,$}}
\newcommand\gsim{\buildrel > \over \sim}
\newcommand{\tnm}{\tablenotemark}
\newcommand{\tnt}{\tablenotetext}

\newcommand{\Msun}{\mbox{$M_{\odot}\;$}}

\begin{document}

\title{MAGNETIC AND SPIN HISTORY OF VERY YOUNG PULSARS}
\author{Alexander Muslimov
        \footnote{Alexander von Humboldt Fellow}}

\affil{Institut f\"{u}r Astrophysik der Universit\"{a}t Bonn,
       Auf dem H\"{u}gel 71, 53121 Bonn, Germany \\
       muslimov@astro.uni-bonn.de}
\and
\author{Dany Page}
\affil{Instituto de Astronom\'{\i}a, UNAM,
       Apdo Postal 70-264,
       04510 M\'{e}xico D.F., M\'{e}xico. \\
       page@astroscu.unam.mx}

\begin{abstract}

  After Michel (1994) introduced a phenomenological
  picture of ``rapid magnetization'' of newly born neutron stars
  (NSs), Muslimov \& Page (1995) suggested
  that the physical conditions accompanying the formation of a NS
  (convection in the NS envelope, high-rate accretion in a supernova)
  may result in the large scale surface magnetic field of the NS
  having a low value ($\sim 10^8-10^9$ G), while the bulk of the magnetic
  flux is submerged under the stellar surface. The subsequent growth
  of the large scale surface magnetic field then occurs owing to ohmic
  diffusion of a strong internal field ($\sim 10^{12}-10^{13}$ G) and
  proceeds over a timescale of $10^2-10^3$ yr, depending on the early
  thermal history of the NS, initial distribution of magnetic flux,
  and electrical resistivity in the crust.  Referring to this
  suggestion, we perform numerical computations to demonstrate its
  possible relevance to young radio pulsars. In our calculations, we
  use different NS structures based on a model of dense matter
  presenting a phase transition to kaon condensation which softens the
  equation of state.  This model implies both slow and fast neutrino
  cooling, depending on the stellar mass (which is assumed to be in the
  range of $1.35-1.50~M_{\odot}$).  We present sequences of secular
  evolution of the surface magnetic field, spin-down luminosity, spin
  period, ``braking index,'' and spin-down age best matching the
  corresponding measured and derived quantities for PSR 0531+21
  (Crab), PSR 0540-69, and PSR 1509-58. We conclude that the effect
  under consideration reproduces remarkably well both the observed
  rotational characteristics and derived values of the surface
  magnetic field strength for these radio pulsars, the only ones
  having reliable measurements of the braking indices.  In addition,
  our analysis suggests that NSs in the Crab and PSR 0540-69 have
  experienced fast neutrino cooling and that their masses are above
  the critical mass for the phase transition while the NS in PSR
  1509-58 has a mass below this critical mass and has therefore
  undergone slow neutrino cooling.

\end{abstract}

\keywords{\quad pulsars: individual
                         (PSR 0531+21, PSR 0540-69, PSR 1509-58) \quad ---
          \quad stars:  neutron \quad ---
          \quad stars: magnetic fields}


\newpage
\section{INTRODUCTION
         \label{sec:1}}

The timing of radio pulsars reveals that their ``clock mechanisms''
are subject to a systematic delay (spinning down) and that various
irregularities in their run such as ``glitches,'' ``microglitches,''
etc., may occur. The measurements of the first and second time
derivatives of pulsars' spin period, $\dot {P}$ and $\ddot {P}$,
respectively, can provide us with invaluable information about the
rotational dynamics of neutron stars (NSs). For the standard formula
describing the secular decrease in the angular velocity $\Omega$ of an
NS,
\be
\dot {\Omega} \propto - \Omega ^n \;\; ,
\label{eq:1}
\ee
where $n=3$ corresponds to the mechanism of pure magnetic dipole
braking, the ``braking index'' (see Manchester \& Taylor
\markcite{MT77} 1977, hereafter MT, eq. [6-17]),
\be
n\equiv {{\Omega \ddot{\Omega }}\over ({\dot{\Omega }})^2} \;\; ,
\label{eq:2}
\ee
can, in principle, be determined from observations. In practice, the
measurements of braking indices are contaminated by the effects of
``restless'' behavior of $\dot {P}$ over short timescales and are a
point of controversy. Only three young pulsars have measured values of
$n$ not dominated by timing noise: PSR 0531+21 (Crab), $n=2.509\pm
0.001$ (Lyne, Pritchard, \& Smith \markcite{LPS88} 1988); PSR 0540-69,
$n=2.01\pm 0.02$ (Nagase et al. \markcite{N90}1990; Gouiffes, Finley,
\& \"{O}gelman \markcite{GFO92} 1992); and PSR 1509-58, $n=2.837\pm
0.001$ (Kaspi et al. \markcite{K94}1994).  The observed rotational
properties of these three pulsars are listed in Table~\ref{tab:1}.

Theoretically, as is summarized in \markcite{MT77} MT (see Table 9-1),
there are a number of factors affecting the braking index,
e.g., multipole electromagnetic radiation ($n\geq 5$), gravitational
quadrupole radiation ($n=5$), magnetic field decay ($n > 3$), radial
deformation of field lines ($1\leq n \leq 3$), pulsar wind ($n < 3$),
relaxation of the NS equilibrium form ($n < 3$), pulsar transverse
velocity ($n < 3$), etc.  To these, one should add two mechanisms
recently proposed: spin-up of the NS during its early evolution owing to
the intense neutrino emission ($n<0$)(Alpar \& \"{O}gelman
\markcite{AO90} 1990) and crustal ``plate'' motions (Ruderman
\markcite{R91} 1991), the latter being able to produce values of $n$
larger or smaller than 3.  Also, Blandford \& Romani \markcite{BR88}
(1988) have discussed the possibility that the braking indices of PSR
0531+21 and PSR 1509-58 may be consistent with a secular growth of
their magnetic field.  They appealed to the thermomagnetic instability
in the NS crust (Blandford, Applegate, \& Hernquist \markcite{BAH83}
1983, see also Urpin, Levshakov, \& Yakovlev \markcite{ULY86} 1986) as
a basic reason for the magnetic field growth.  Finally, Michel
\markcite{M91} (1991) has also emphasized that the monotonic increase
of the magnetic field with time would give braking indices less than 3
in all young pulsars where it can be determined.

In this paper we employ our recent calculations of growth of the
surface magnetic field in a newly born NS (Muslimov \& Page
\markcite{MP95} 1995, hereafter MP) to demonstrate the striking
agreement of our model with the available observational data for the
Crab, PSR 0540-69, and PSR 1509-58. Our tentative analysis (see
\markcite{MP95} MP) had been stimulated by a phenomenological picture
recently discussed by Michel \markcite{M94} (1994, hereafter M94) that
the magnetic field of a NS at birth is low and increases, by a factor
of $\gsim 10^3$, to the typical value of $\sim 10^{12}$ G during a
period of a few hundred years following the supernova.
\markcite{MP95} MP suggested, as a physical basis for this picture,
that the surface magnetic field of a young NS may grow from a very low
value ($\sim 10^8-10^9$ G) owing to ohmic diffusion of a strong
internal magnetic field ($\gsim 10^{12}$ G) that was initially
submerged under the surface layers.  Assuming that an NS spins down as
a result of magnetic dipole radiation, we determine the free fitting
parameters of our model. For example, for each of these pulsars we
specify the initial spin period and initial profile of the magnetic
flux in the crust.

These early magnetic and spin evolutions are strongly affected by the
thermal evolution of the NS. In the present study we exploit different
stellar structures and thermal histories to fit the data for the three
young pulsars. Specifically, we consider different NS masses and the
kaon condensation phase transition (Thorsson, Prakash \& Lattimer
\markcite{TPL94} 1994; hereafter TPL) in the NS core which results in
different cooling histories depending on the mass.  Within the
framework of our analysis, the fast cooling of an NS provides the best
fit for the Crab and PSR 0540-69, including accounting for the
discrepancy between their ``real'' and spin-down ages. In the case of
PSR 1509-58, we will need a slow cooling, i.e., a star of mass below
the critical mass for the phase transition.

In $\S~\ref{sec:2}$ we present our numerical model, and in
$\S~\ref{sec:3}$ we illustrate matching of our calculations with the
observed parameters for PSR 0531+21 (Crab), PSR 0540-69, and PSR
1509-58. Our principal conclusions are summarized in $\S~\ref{sec:4}$.

\section{BASIC ASSUMPTIONS AND METHOD
         \label{sec:2}          }

\subsection{General Picture
            \label{sec:2.1}}

We assume that an NS is born with a spin period in the range 20-40 ms.
Alternatively (see, e.g., \markcite{M94} M94), the luminosities of the
NSs born as rapid rotators will become large before the NSs slow down
to the present luminosities of the Crab, PSR 0540-69, and PSR 1509-58
which are already large enough. To calculate the spin evolution of the
NS we use the standard formula (see, e.g., \markcite{MT77} MT, p. 110,
180) for the spin-down torque resulting from emission of magnetic dipole
radiation.

If the secular evolution of the dipole magnetic moment of an NS is the
only physical reason affecting its spinning down, it is
straightforward to derive the following expression for the braking
index
\be
n=3+2 \left( {{{\dot{B}}_{surf}} \over {B_{surf}}} \right)
\left( {\Omega \over {\dot{\Omega }}} \right) \;\; ,
\label{eq:3}
\ee
where $B_{surf}$ is the surface value of the magnetic field strength
at the magnetic pole.  (A similar formula, but missing a factor 2 in
the second term, has been used by Chanmugam \& Sang \markcite{CS89}
[1989] in their analysis of the possible response of braking indices
to ohmic field decay in pulsars.)

In formula (\ref{eq:3}) $\dot{\Omega }$ is always negative and,
depending on whether ${\dot{B}}_{surf}>0$ (field growth) or
${\dot{B}}_{surf}<0$ (field decay), we have $n<3$ or $n>3$,
respectively. Here we attempt to quantitatively examine the
possibility that, in a newly born NS, the magnetic field (of strength
$\sim 10^{12}-10^{13}$ G), initially trapped under the stellar
surface, diffuses ohmically up through the crust (\markcite{MP95} MP).
The main uncertainty in this picture is the form of the initial
distribution of the magnetic flux in the upper crust. For illustrative
purposes, we describe the initial profile of magnetic flux in the crust
by referring to the depth down to which the flux is submerged under
the stellar surface. Approximately at this depth the magnetic flux has
a maximum value and decreases by a factor of $\sim 10^2-10^3$ toward
the stellar surface.

\subsection{Basic Equations and Solution
            \label{sec:2.2}}

We now consider the evolution of the dipole component of a purely
poloidal magnetic field $\bf B$ in the NS crust. For a dipole mode, in
spherical coordinates ($r$, $\theta $, $\phi $), we can express $\bf
B$ in the form
\be
{\bf B}=B_0 \left(
{S\over {r^2}} \cos \theta \; {\bf e}_r -
{1\over {2r}} {{\partial S}\over {\partial r}} \sin \theta \; {\bf
e}_{\theta }
\right) \;\;,
\label{eq:4}
\ee
where $B_0$ is some normalization magnitude of the field, ${\bf e}_r$
and ${\bf e}_{\theta }$ are unit vectors in the radial and meridional
directions, and $S = S(r,t)$ is the stream function.

Using the standard equation for the magnetic field evolution in flat
spacetime (neglecting the hydrodynamic motions, thermomagnetic
effects, and anisotropy of the electrical conductivity),
\be
{{\partial {\bf B}}\over {\partial t}}=-{{c^2} \over {4 \pi }}
\nabla \times \left( {1 \over {\sigma }} \nabla \times {\bf B} \right)
\;\; ,
\label{eq:5}
\ee
we arrive at the following equation for the evolution of $S$:
\be
{{\partial S}\over {\partial t}}={{c^2} \over {4\pi \sigma }}
\left( {{\partial ^2 S}\over {\partial r^2}}-{{2S} \over {r^2}}
\right) \;\; ,
\label{eq:6}
\ee
where $\sigma = \sigma (r,t)$ is the electrical conductivity of
matter.

We solve equation (\ref{eq:6}) numerically subject to the appropriate
boundary conditions at the stellar surface and at the base of the
crust. At the surface ($r=R$) we impose the standard boundary
condition that the internal field merges continuously with an external
vacuum field. For a dipole field this condition reads
\be
R{{\partial S} \over {\partial r}} + S = 0 \;\; .
\label{eq:7}
\ee
At the base of the crust ($r=r_b$) we assume that
\be
S= \rm Const \;\; .
\label{eq:8}
\ee
This second condition is perfectly justified for our problem, and it
means that the bulk of the magnetic flux [the magnetic flux through a
hemisphere of radius $r$ is proportional to the value of the function
$S(r,t)$] is frozen into the inner crust (outer core), at least over a
timescale of $\gsim 10^6$ yr. The ohmic diffusion and decay of the
magnetic field in the innermost layers of the crust occur on a much
longer timescale and are of no importance for our present purpose.
Note that in our calculations the conductivity of the crust depends on
both the density and temperature of matter (see, e.g., \markcite{MP95}
MP and references therein for a relevant discussion of the
conductivity regimes in the crust and corresponding numerical
estimates), the latter changing with time as the NS cools down.

We perform our calculations of the spin-down luminosity, spin period,
braking index, and spin-down age ($\tau_{sd} = P/2{\dot{P}}$) of the
NS by solving self-consistently the system of equations describing
both its magnetic and spin evolution (owing to magnetic dipole
braking).  We also employ formula (\ref{eq:3}) in our field evolution
code to calculate the secular evolution of the braking indices.

The principal parameters needed to be specified in our computations
are the initial spin period of the NS, $P_0$, and the initial surface
value of the magnetic field strength, $B_{surf}^0$. We assume that for
the Crab, PSR 0540-69, and PSR 1509-58 the value of $B_{surf}^0$ is in
the range of $(1-5)\times 10^{10}$ G, while $P_0$ ranges from $\sim
20$ to $\sim 40$ ms. It is important that our choice of $P_0$ for the
Crab and PSR 0540-69 is not arbitrary, and we shall discuss this
below.

Our calculations show clearly that the character of early magnetic and
spin evolutions of a NS is determined by the dynamics of freezing and
the structure of its crust. For example, the stellar models allowing
for a rapid cooling result in a power-law temporal growth of the
surface value of the magnetic field strength in the saturation regime.
In contrast, the field evolution in the saturation regime for a
relatively slowly cooling NS is more complex and cannot be described
by a simple power-law time dependence.  Also, the fact that the NSs in
the Crab and probably in PSR 0540-69 have spin-down ages well
exceeding their real ages can be understood naturally within the
framework of our analysis only if we assume that these NSs have
undergone a stage of rapid neutrino cooling. The corresponding stellar
models and thermal histories we have employed in our calculations for
these pulsars do reproduce remarkably well the discrepancy between
their spin-down ages $\tau _{sd}$ and expected real ages $t_{model}$
(see $\S~\ref{sec:3}$). Finally, in the case of PSR 1509-58 our
results may testify that the spin-down age of the associated NS is
less than its real age and that this can be attributed to relatively
slow neutrino cooling of the NS.

The standard equation describing the evolution of the spin period
owing to magnetic dipole losses can be easily integrated to give
\be
P=P_0 \left( 1+ \alpha \int _{0}^{t} B_{surf}^2 dt' \right) ^{1/2}
 \;\; ,
\label{eq:9}
\ee
where $\alpha $ is a numerical factor depending on the
radius, moment of inertia, and initial value of the spin period of the NS.
Using the above expression, we can represent the spin-down age of an NS
in the form
\be
\tau _{sd}\equiv {P \over {2 \dot {P}}} =
{K \over {B_{surf}^2}} \int _{0}^{t} B_{surf}^2 dt' \;\; ,
\label{eq:10}
\ee
where $K=[1-(P_0/P)^2]^{-1}$.  Note that in our problem the integral
$\int _{0}^{t} B_{surf}^2 dt'$ is dominated by the contribution from
the saturation regime, and the integration can therefore be performed
over the interval of time during which the field has been saturating.
As we have emphasized above, for the stellar models with a rapid
neutrino cooling (that have used for NSs in the Crab and PSR
0540-69) the surface value of the magnetic field strength increases in
the saturation regime according to a power low. Thus, assuming that in
the saturation regime, for NSs in the Crab and PSR 0540-69, $B_{surf}
\propto t^{\epsilon }$, we obtain
\be
\tau _{sd} \sim  {K \over {2 \epsilon +1}} t
\label{eq:11}
\ee
and
\be
n \sim 3-{{2 \epsilon } \over {2 \epsilon +1}}K \;\; .
\label{eq:12}
\ee
{}From equations (\ref{eq:11}) and (\ref{eq:12}) we find that
\be
K \sim 3-n+{{\tau _{sd}} \over t}
\label{eq:13}
\ee
and
\be
\epsilon \sim {{3-n} \over {2(n-3+K)}} \;\; .
\label{eq:14}
\ee

We can use the expressions (\ref{eq:13}) and (\ref{eq:14}) for rough
estimates of the values of $P_0$ and $\epsilon $ for the Crab and PSR
0540-69.  Substituting the values $P$, $n$, $\tau _{sd}$, and
$t=t_{model}$ into equations (13) and (14), we find that for these
pulsars $P_0 \sim 23$ ms and $\epsilon \sim 0.18$, and $P_0 \sim 41$
ms and $\epsilon \sim 0.27$, respectively.  These estimates justify
our choice of the range of the initial spin periods for the Crab and
PSR 0540-69. However, the above reasoning is not applicable to the NS
in PSR 1509-58 because of the needed relatively slow stellar cooling
and, as a consequence of this, more complex temporal evolution of the
surface value of its field strength in the saturation regime. In this
sense, our choice of $P_0$ for PSR 1509-58 is quite arbitrary.

For a given range of variation of $P_0$ and $B_{surf}^0$, we survey
different profiles of the initial distribution of the magnetic flux in
the crust to fit the observed and derived parameters such as $P$, $n$,
$\tau _{sd}$, and $B_{surf}$. As a result of such a fitting we are
able to determine the sensible range of the initial depths of
submergence ($z_{sub}$) of the magnetic field in the crust (see
Table~\ref{tab:1}).  Note that the rate of spinning down and the value
of the braking index decrease with increasing the initial depth of
submergence $z_{sub}$ of the flux (for a given NS structure and
thermal evolution), with the values of braking indices being most
sensitive to the variation of this depth as a parameter. In our
calculations, the initial profile and depth of submergence of the
magnetic flux in the crust are therefore determined by matching the
theoretical and measured values of braking indices for a given
(derived) spin-down age of the pulsar.  Note also that the variation
of $P_0$ only slightly affects the calculated values of the spin
period and braking index for relatively long-period pulsars such as
PSR 1509-58.

\subsection{Neutron Star Models and Cooling
            \label{sec:2.3}}

Our many trials to fit the observed parameters of the three young pulsars
led us to focus on different NS thermal histories and
stellar structures. To perform a consistent analysis, we are thus forced
to assume that the three NSs have different masses and that the equation of
state (EOS) allows fast or slow neutrino cooling depending on the mass.

A number of emission mechanisms have been proposed for the fast
neutrino cooling (see, e.g., Pethick \markcite{P92} 1992, Page
\markcite{P94} 1994). In this paper, we adopt the possibility of kaon
condensation which is presently the scenario most seriously discussed
in the literature (see Brown \& Rho \markcite{BR95} 1995 for a
review). To build a consistent EOS, we follow the method of TPL, in
which the effect of kaon condensation is superposed on an underlying
purely baryonic EOS in the mean field approximation (very similar
results have been obtained with a different formalism by the Kyoto
group [see, e.g., Tatsumi \markcite{T95} 1995]). Due to the large
uncertainties in the structure of matter at the densities
characteristic of NS cores, we have many free choices. We select the
baryonic EOS in a standard way, while we parameterize the kaon
condensation properties to satisfy our needs, namely:

$1.$ We select, as TPL did, the baryonic EOS from the set of parametric
models of Prakash, Ainsworth \& Lattimer \markcite{PAL88} (1988,
hereafter PAL): we use the model ``PAL33'' with an incompressibility
$K_0 = 240$ MeV and a symmetry energy function $F(u) = u^{1/2}$ (where
$u=n/n_0$, $n$ is the baryon number density and $n_0$ = 0.16 fm$^{-3}$
is the ``saturation'' density).  This EOS is intermediate in stiffness
between the standard microscopic non relativistic FP EOS (Friedman \&
Pandharipande \markcite{FP81} 1981) and the microscopic relativistic
MPA EOS (M\"uther, Prakash \& Ainsworth \markcite{MPA87} 1987). It
provides a proton fraction that is low enough to prevent the fast
neutrino emission by the direct Urca process (Lattimer et al.
\markcite{LPPH91} 1991).  The mass-radius relationships for these
three EOS are shown in Figure~\ref{fig:1}.

$2.$ The properties of the kaon condensate in the formalism developed
by TPL depend strongly on the values of the coupling constants of the
model Lagrangian, all of them but one being fortunately constrained by
experiments.  The only poorly constrained parameter is thus $a_3$
(expressed as $a_3 m_s$, where $m_s$ is the mass of the strange
quark), and its reasonable range of variation is from $-134$ MeV down
to about $-300$ MeV (see TPL). Given the underlying baryonic EOS, the
value of $a_3 m_s$ determines the critical density for the
condensation phase transition, i.e., the critical mass for an NS to
contain a kaon condensate core.  Note that the maximum NS mass also
depends on $a_3 m_s$ but not so sensitively as the critical mass. We
adopt in our calculations the value of $a_3 m_s = -190$ MeV. The
resulting mass-radius relationship and the density profiles for five
models of different masses are plotted in Figures~\ref{fig:1} and
\ref{fig:2}. The NSs with masses below 1.37 \Msol, the critical mass
for appearance of the condensate, are ``standard'' NSs, while more
massive ones should rather be called ``nuclear'' stars (Brown \& Bethe
\markcite{BB94} 1994), since they contain roughly equal amounts of
protons and neutrons, as laboratory nuclei do, and a very strong kaon
condensate (unlike laboratory nuclei).

The motivation for our particular choice of $a_3 m_s = -190$ MeV is
twofold: first, it provides a transition from the slow to the fast
neutrino cooling at a mass of 1.37 \Msol , which is within the range
of measured NS masses (van Kerkwijk, van Paradijs \& Zuiderwijk
\markcite{KPZ95} {1995).
Second, it gives a maximum mass of 1.575 \Msol that is still
consistent with the data and is also compatible with the possibility
that the NS formed in SN 1987A went into a black hole (Bethe \&
Brown \markcite{BB95} 1995).

With regard to the cooling we come up with the following scenario.  A
NS of mass $M <$ 1.37 \Msol allows only the slow modified Urca process
along with the associated nucleon bremsstrahlung processes and
therefore follows the ``standard'' cooling scenario .  If $M >$ 1.37
\Msol , then the presence of the kaon condensate increases the
neutrino emission by orders of magnitude (Brown et al.
\markcite{BKPP88} 1988) resulting in fast cooling (Page \& Baron
\markcite{PB90} 1990). Above 1.463 \Msol the direct Urca process is
also allowed and dominates the total neutrino emissivity that results
in an even faster cooling (Page \& Applegate \markcite{PA92} 1992).
The cooling curves for five models are shown in Figure~\ref{fig:3},
where the occurrence of the three different scenarios is clearly seen.
The evolution of the profiles of internal temperature for two of these
models is shown in Figure~\ref{fig:4}.  The profiles have strong
gradients during the early phases of cooling, but after a few years
the crust temperature in the density range 10$^{11}$ - 10$^{14}$
g$\cdot $cm$^{-3}$, where the currents producing the magnetic field
are located, is uniform to within 50\% or better.

Some details of the method used to produce these cooling models can be
found in Page \& Baron \markcite{PB90} (1990) and Page \& Applegate
\markcite{PA92} (1992). In this paper we introduce some modifications,
which, besides the changes in the EOS, are as follows. We use a proper
treatment of the suppression of the specific heat and neutrino
emission by baryon pairing according to Levenfish \& Yakovlev
\markcite{LY94a,LY94b} (1994a,b).  For the kaon condensate neutrino
emissivity we use the formula
\be
\epsilon_{\nu}^{K} = 7.6 \times 10^{25}
                     \times T_9^6
                     \left(\frac{m_n^*}{m_n}\right)^2
                     \left(\frac{\mu_e}{m_{\pi}}\right)
                     \frac{\sin^2 \theta_K}{2}
                     [1+3(g_A')^2] \sin^2 \theta_C
                     ~ \rm erg \; cm^{-3} \; s^{-1} ,
\label{eq:15}
\ee
which is obtained from the similar formula of Brown et al.
\markcite{BKPP88} (1988) by replacing the $\theta_K^2$ by $\sin^2
\theta_K$ to take into account that $\theta_K$ is not small. Also, an
additional factor of 2 has been included following Thorsson et al.
\markcite{TPTP95} (1995). In the above formula $m_n$ and $m_n^*$ are,
respectively, the neutron free and effective masses; $\mu_e$ is the
electron chemical potential; $m_{\pi}$ is the pion mass; $g_A' \sim 1$
is the in-medium Gamow-Teller coupling constant; $\theta_C = 13^\circ$
is the Cabbibo angle; $T_9$ is the temperature in 10$^9$ K; and
$\theta_K$ is the chiral angle characterizing the kaon condensate
strength ($\theta_K$ grows rapidly to near 90$^\circ$ above the
critical density).  We have incorporated nucleon pairing by employing
the results of, correspondingly, Ainsworth, Wambach, \& Pines
\markcite{AWP89} (1989) [for neutron $^1$S$_0$ pairing (inner crust)],
Takatsuka \markcite{T72} (1972) [for neutron $^3$P$_2$ pairing
(core)], and Takatuska \markcite{T73} (1973) [for proton $^1$S$_0$
pairing (core)]. These core neutron and proton pairings (see Page
\markcite{P94} 1994 for a comparative presentation of the relevant
calculations) do not affect our computations because at the densities
typical for the inner core (where the intense neutrino emission
occurs) the corresponding critical temperatures $T_c$ are vanishingly
low and do not allow the pairings. In contrast, neutron pairing in the
inner crust is more important, since it reduces substantially the
specific heat in this region, where the main currents producing the
stellar magnetic field are located. As a result, the cooling of the
crust proceeds faster during the first few decades. Note that the
calculation of Ainsworth et al. \markcite{AWP89} (1989) is the most
accurate treatment available to date and gives $T_c$ of the order of
10$^{10}$ K in most of the inner crust. The uncertainty in the profile
of $T_c$ does not affect our results (as can be seen from
Fig.~\ref{fig:4}), since the temperature of the inner crust drops
below 10$^{10}$ K very rapidly (within a timescale much shorter than
that for the magnetic field evolution in the inner crust).

\section{RESULTS OF NUMERICAL CALCULATIONS AND DISCUSSION
         \label{sec:3}                             }

The three observables are $P$, $\dot{P}$, and $\ddot{P}$, or
equivalently $P$, $\tau_{sd}$ (and $B_{surf}$ with redundancy), and
$n$.  These observables are listed in Table~\ref{tab:1} together with
the parameters of our models.  Our models aim at fitting these three
(four with $B_{surf}$) observed values at a given model age
$t_{model}$ as listed in Table~\ref{tab:1}.  The age of the Crab
pulsar is known, and the $t_{model}$ is taken to be equal to this age.
In the case of PSR 0540-69, the study of the associated supernova
remnant (SNR 0540-69) indicates an age $\tau_{SNR}\sim 760$ yr
(Kirshner et al.  \markcite{KMWB89} 1989), but the actual age may be
somewhat larger or smaller depending on whether the remnant's
expansion has been accelerated by the pulsar or slowed down by the
interaction with the interstellar medium.  We thus only impose that
$t_{model} \sim \tau_{SNR}$.  The age of PSR 1509-58 is highly
uncertain, and $t_{model} \sim$ 1,860 yr is an output of our fit for
this pulsar. If one accepts the association of PSR 1509-58 with the
supernova SN 185 (Thorsett \markcite{T92} 1992), then this age is very
close to the real age (however, Strom \markcite{S94} [1994] criticized
this association).

We show in Figure~\ref{fig:5} the temporal evolution of the surface
magnetic field $B_{surf}$ (strictly speaking, its component
perpendicular to the spin axis), spin-down luminosity $L_{sd}$, spin
period $P$, braking index $n$, and spin-down age $\tau_{sd}$ for three
different models.  The present positions (i.e., at age $t_{model}$) on
the evolutionary tracks are marked by dots. Figures 5A, 5B, and 5C
show clearly that all observable and derived (such as $B_{surf}$)
quantities for the Crab, PSR 0540-69, and PSR 1509-58 can be
reproduced within our model. Note that the calculated values of
$B_{surf}$ exactly match the corresponding values derived from the
standard formula for the magnetic dipole losses in which the stellar
radius and moment of inertia are taken to be equal to those the NS
model used for each particular pulsar.

\subsection{Effects of Cooling and Stellar Structure
            \label{sec:3.2}}

Figure~\ref{fig:5}E shows that with the fast neutrino cooling
$\tau_{sd}$ exceeds $t_{model}$ for a longer time than in the case of
slow cooling, but the detailed behavior depends also on other
quantities such as the $n$ and $t_{model}$.  The requirement that the
model fits the age of the pulsar (besides $P$, $\tau_{sd}$, and $n$)
thus puts strong constraints on the thermal evolution of the NS. In
the cases of the Crab and PSR 0540-69, such a fit can only be achieved
by employing fast neutrino cooling, which results in a rapid freezing
of the stellar crust and corresponding decrease (see also
\markcite{MP95} MP) in the electrical resistivity of the crust.  This
is clearly seen in Figure~\ref{fig:5}A: after $t\sim 20$ yr, the rate
of growth of the surface magnetic field slows down substantially, and
the magnetic field enters the saturation regime.  The Crab pulsar has
been claimed several times (see, e.g., Tsuruta \markcite{T86} 1986;
Page \& Baron \markcite{PB90} 1990) to be a typical example of an NS
that follows the standard cooling scenario.  However, there is
actually no real observational evidence in favor of these claims (Page
\markcite{P94} 1994, \S 6.2) and this pulsar may perfectly well have
undergone fast neutrino cooling as we need here.  In the case of PSR
0540-69, the progenitor mass is estimated to be around 15-20 \Msun
(Kirshner et al.  \markcite{KMWB89} 1989). The latter implies a high
mass NS, for which a fast cooling (needed in our analysis to fit the
data) is almost inevitable. In the case of PSR 1509-58 there is no
such age constraint.  However, if our model is correct, then the fit
of $P$, $\tau_{sd}$, and $n$ is evidence in favor of slow neutrino
cooling, which, in its turn, would argue for a progenitor less massive
than that of the Crab [which is estimated to be about 9 \Msol (see,
e.g., Nomoto \markcite{N85} 1985)].

Our conclusion concerning the cooling, fast for the Crab and PSR
0540-69 and slow for PSR 1509-58, is quite robust within the framework
of our study. However, our analysis cannot discriminate between
different fast cooling scenarios owing, e.g., to the kaon or pion
condensates or direct Urca from nucleons and/or hyperons. Note also
that our calculations are less sensitive to the stellar radius, and
the complexity of internal field (the presence of higher order
multipoles, toroidal component, etc.) may also allow us to relax
somewhat the constraints on the crust thickness.

\subsection{Predictions for the Crab and SN 1987A
\label{sec:3.3}}

The detection of thermal emission from the surface of the NS in the
Crab can, in principle, establish the occurrence of the fast neutrino
cooling for this star in the past.  The new {\em ROSAT} upper limit
(Becker \& Ashenbach \markcite{BA95} 1995), based on the absence of
detection, is already marginally inconsistent with the standard
cooling scenarios.  Our cooling scenario implies an effective
temperature well below that in the standard scenario which may
potentially be discriminated by the future X-ray observatories.
Another interesting point is that, as one can see in
Figure~\ref{fig:5}B, our calculations imply that the spin-down
luminosity of the Crab has been decreasing only slightly during the
past centuries. The comparison of the present remnant luminosity with
earlier observations may confirm or rule out this possibility. We have
used an NS model of mass 1.45 \Msol for the Crab which is required by
the need for fast cooling within our TPL EOS.  However, this
precise value of mass should not be taken too seriously, since the
critical mass for the kaon condensation is model dependent, and the
value of $a_3 m_s$ we have adopted in our calculations is arbitrary.
Owing to the estimated low mass of the Crab progenitor (Nomoto
\markcite{N85} 1985) one may suggest a lower mass for the associated
NS. In this case, in terms of NS modeling, we would simply need to
increase $|a_3 m_s|$ to reduce the critical mass for kaon
condensation, without affecting seriously the maximum mass (which can
also be adjusted by changing the underlying baryonic EOS). Thus, our
estimate of the NS mass for the Crab is not warranted.

The development of our model was triggered by the announcement of the
discovery of an optical pulsar in the remnant of SN 1987A. This
detection has been neither confirmed nor ruled out by Hubble Space
Telescope observations (Percival et al. \markcite{P95} 1995). Our
original implication (\markcite{MP95} MP) of the calculations of the
field growth for the possible NS in the SN 1987A was based on a
stellar model with slow neutrino cooling and resulted in a slow
increase of the pulsar's spin-down luminosity, similar to our present
model for PSR 1509-58.  This first scenario may explain the lack of
evidence of powering of the remnant by the pulsar and predicts that
this powering should be detectable in a few decades or centuries.
However, our modeling of PSR 0540-58, whose progenitor seems to be a
twin of the progenitor of SN 1987A (Kirshner {\em et al.}
\markcite{KMWB89} 1989), argues for fast neutrino cooling in the case
of NS 1987A. If this very young NS follows similar evolution as that
in PSR 0540-58, its spin-down luminosity should already be quite high,
and it will be growing up by an order of magnitude during the next
decade or so and may become detectable.  If, however, the magnetic
field of the NS 1987A has been submerged down to deeper layers than in
the case of PSR 0540-58, then the field growth will be retarded, and
the pulsar will be hidden within the remnant for a much longer time.
The other possibility is, of course, that the NS 1987A did go into a
black hole (Brown, Bruenn, \& Wheeler \markcite{BBW92} 1992). This
scenario puts strong constraints on the maximum NS mass , $\sim
1.56$~\Msol (Bethe \& Brown \markcite{BB95} 1995), to which our TPL
EOS has been adjusted. Also, the scenario of Brown et al.
\markcite{BBW92} (1992) stipulates that the mass of the NS 1987A only
slightly exceeded this maximum mass. The fact that the NS in PSR
0540-58 (a probable twin of the NS 1987A) did survive both reinforces
their suggestion and may imply that it has a mass just below the
maximum mass and has undergone a phase of fast neutrino cooling.

\section{SUMMARY AND CONCLUSIONS
         \label{sec:4}}

         We have presented numerical calculations of early magnetic
and spin evolutions of an NS born with a magnetic flux initially
trapped under the stellar surface. We have demonstrated explicitly
that our calculations may be relevant to the magnetic and spin
evolutions of the NSs in PSR 0531+21 (Crab), PSR 0540-69, and PSR
1509-58.

         Our principal findings are as follows.

\indent{1. The braking indices of the Crab pulsar ($n=2.5$),
  PSR 0540-69 ($n=2.0$), and PSR 1509-58 ($n=2.8$) can be
  understood easily in terms of the standard mechanism of
  magnetic dipole braking, provided that the large scale surface magnetic
  field of the associated NSs has been growing up with the
  rate corresponding to the ohmic diffusion of an internal
  large scale magnetic field, initially submerged under the stellar
  surface.}

\indent{2. We account for the existing difference ($\sim $ 300 yr)
  between the spin-down and real ages for the Crab pulsar. Also,
  according to our calculations, the real age of PSR 0540-69 is about
  890 yr, i.e., 770 yr less than its spin-down age, in agreement with
  the estimated age of the associated SNR 0540-69.  Finally, the real
  age of PSR 1509-58, in contrast to the Crab and PSR
  0540-69, may exceed its spin-down age by about 300 yr.}

\indent{3. In our model the NSs in the
  Crab, PSR 0540-69, and PSR 1509-58 have initial spin periods of,
  respectively, $P_0=21$, 39, and 20 ms, and their initial
  surface magnetic fields are in the range of $(1-5)\times 10^{10}$ G.}

\indent{4. The spin-down luminosity of the Crab, PSR 0540-69,
  and PSR 1509-58 reaches its maximum at an NS age of $\sim 140$,
  $\sim 250$, and $\sim 80$ yr, respectively.}

\indent{5. Our analysis may sensibly indicate that NSs in the Crab and
  PSR 0540-69 have thin ($\lsim 200$ m) outer crusts and have
  experienced a phase of fast neutrino cooling. In contrast, the NS in
  PSR 1509-58 probably has a relatively thick ($\sim 430$ m) outer
  crust, and has undergone a slow neutrino cooling.}

To extend our investigation of the early magnetic and spin evolution
of an NS to other young pulsars, we need to have reliable measurements
of their braking indices that are not available at the moment.
However, the observations recently initiated with the {\it ROSAT}
X-ray telescope (see, e.g., Aschenbach, Egger, \& Tr{\"{u}}mper 1995) of
fine structures in young supernova remnants (SNRs) containing pulsars
may provide us with more accurate estimates of the real ages of young
radio pulsars associated with the SNRs. Given such estimates, we would
be able to proceed farther with a more extensive analysis of the
effect under consideration in pulsars.

Finally, a more detailed study should incorporate some other possible
mechanisms mentioned in the introduction. Spin-up of an NS by early
neutrino emission can be incorporated in our model since it is a well
defined mechanism and requires the same ingredients of stellar
structure and cooling. The various mechanisms reducing $n$ below 3 are
very model dependent and would increase the number of free parameters.
Quadrupolar radiation, magnetic and/or gravitational (resulting in $n
> 3$) can reasonably be excluded. The crustal ``plate'' motion is of
major uncertainty and is practically impossible to model
quantitatively. Although plate tectonics is hypothetical, crust
cracking (``plate tectonic'' activity ?) seems to be taking place in
the Crab (Link, Epstein, \& Baym 1992; Alpar et al. 1994), and crustal
plates may manifest themselves at the surface of the Geminga NS (Page,
Shibanov, \& Zavlin 1995).

\acknowledgments 

A.M. thanks  the Alexander-von-Humboldt-Stiftung for  support. At UNAM,
this work has been supported by a UNAM-DGAPA grant IN105794 and a {\it
C\'{a}tedra Patrimonial} from Conacyt. The authors also thank an
anonymous referee for useful comments, which have improved the clarity
of the manuscript.

\clearpage

\begin{small}

\begin{deluxetable}{cccccccccc}
\tablecaption{Characteristics of Pulsars and Parameters of Models \tnm{a}
              \label{tab:1}}
\tablehead{
 PSR     &  $P$  &    $n$            & $\tau_{sd}$  &  $\tau_{SNR}$ &
        $t_{model}$ & $P_0$  & Mass  & $z_{sub}$ & $\rho_{sub}$  \\
         &  (ms) &                   &     (yrs)    &      (yrs)
       &   (yrs)     & (ms )  & $M_{\odot }$
                                    &  (m) & $(10^{10} \rm g/cm^3)$ }
\startdata
0540--69 &  50.4 & 2.01  (2) \tnm{b} &   1660      & $ 760 \pm 100$ \tnm{c}
 &     890     &  39  &  1.50 &   85-90   & 2.62 -- 3.16  \nl
0531+21  &  33.4 & 2.509 (1) \tnm{d} &   1260      & $ 810 \pm 100$ \tnm{c}
 &     940     &  21    &  1.45 &   95-100  & 1.94 -- 2.34  \nl
1509--58 & 150.2 & 2.837 (1) \tnm{e} &   1549      &       1810    \tnm{f}
 &    1860    &  20    &  1.35 &  170-175  & 1.43 -- 1.56  \nl
\enddata

\tnt{a}{Listed are: pulsar spin period $P$, braking index $n$, and spin-down
 age $\tau_{sd}$;
                    associated supernova remnant age $\tau_{SNR}$;
                    model age $t_{model}$ of the NS (as used in our
                    calculations) and initial pulsar spin period $P_0$;
                    theoretical mass of the NS;
                    depth $z_{sub}$ and density $\rho_{sub}$ of initial
                    submergence  of the magnetic field below the surface.}
\tnt{b}{Nagase et al. 1990; Gouiffes et al. 1992.}
\tnt{c}{Kirshner et al. 1989.}
\tnt{d}{Lyne et al. 1988.}
\tnt{e}{Kaspi et al. 1994.}
\tnt{f}{SN 185: Thorsett 1992.}

\end{deluxetable}

\end{small}

\clearpage

\clearpage

\begin{figure}
\caption{Mass-radius relationships for the four NS EOSs:
         FP  (Friedman \& Pandharipande 1981),
         MPA (M\"uther {\em et al.} 1987),
         PAL (Prakash {\em et al.} 1988: see text for the specific
              parameters used), and
         TPL (Thorsson {\em et al.} 1994: see text for the specific
              parameters used).
         The regions in which $M$ decreases with $R$ are unstable, so
         that for the TPL EOS the branch b-c collapses onto the branch
         d-c, and the branch a-b is probably not realized in nature
         and also collapses onto c-d. As a result, for the TPL EOS
         the normal NSs can be produced for gravitational masses below
         1.37 \protect\Msol (point a), while above this mass the leap
         into the regime of formation of nuclear stars (branch c-e with
         kaon condensate) occurs. Point e denotes the maximum mass of
         1.575 \protect\Msol corresponding to this EOS.
\label{fig:1}}
\end{figure}

\begin{figure}
\caption{Density profiles for five neutron-nuclear stars from our TPL EOS.
         At the kaon condensation phase transition the density jumps from
         $\rho_1 = 1.03 \times 10^{15}~ \rm g \; cm^{-3}$ to
         $\rho_2 = 1.76 \times 10^{15}~ \rm g \; cm^{-3}$ and, owing to the
         enormous amount of kaons present, the proton fraction jumps from
         9.85\% at $\rho_1$ up to 54\% at $\rho_2$, while the electron
         fraction drops from 6.14\% down to 0.0022\%.
         Despite the large proton fraction and owing to the very low
         electron fraction the direct Urca process is forbidden at
         densities above $\rho_2$ and is allowed only when $\rho$
         reaches $\rho_3 = 2.8 \times 10^{15}~ \rm g \; cm^{-3}$.
         See TPL for more details.
\label{fig:2}}
\end{figure}

\begin{figure}
\caption{Cooling curves for five stellar models with the TPL EOS.
         The 1.35 \protect\Msol star cools according to the ``standard''
         scenario, the 1.40 and 1.45 \protect\Msol stars undergo fast
         neutrino cooling from the kaon condensate, and the two
         heaviest stars
         allow also the direct Urca process
         (see Fig.~\protect\ref{fig:2}) and cool even faster.
         The ordinate is the redshifted effective temperature.
\label{fig:3}}
\end{figure}

\begin{figure}
\caption{Profiles of the internal temperature for two cooling stars
         with the TPL EOS. The profiles for the 1.40, 1.50, and 1.55 \Msol
         stars are similar to the 1.45 \Msol case.
\label{fig:4}}
\end{figure}

\begin{figure}
\caption{Plot of
         (A) the surface value of the polar magnetic field strength,
         (B) spin-down luminosity,
         (C) spin period,
         (D) braking index, and
         (E) spin-down age
         vs. the model age of the NS.
         Dots on the curves correspond to the ``empirical'' and derived
         data for these pulsars (see \protect\S~\protect\ref{sec:3}
         for details). The parameter value $L_0$ in (B) is equal
         to $2.4\times 10^{34}$, $1.4\times 10^{33}$, and
         $4.2\times 10^{35}$ erg$ \; $s$^{-1}$ for the Crab,
         PSR 0540-69, and PSR 1509-58, respectively. The dependence
         $\tau_{sd}=t_{model}$ is shown by the dotted line in (E).
\label{fig:5}}
\end{figure}

\end {document}
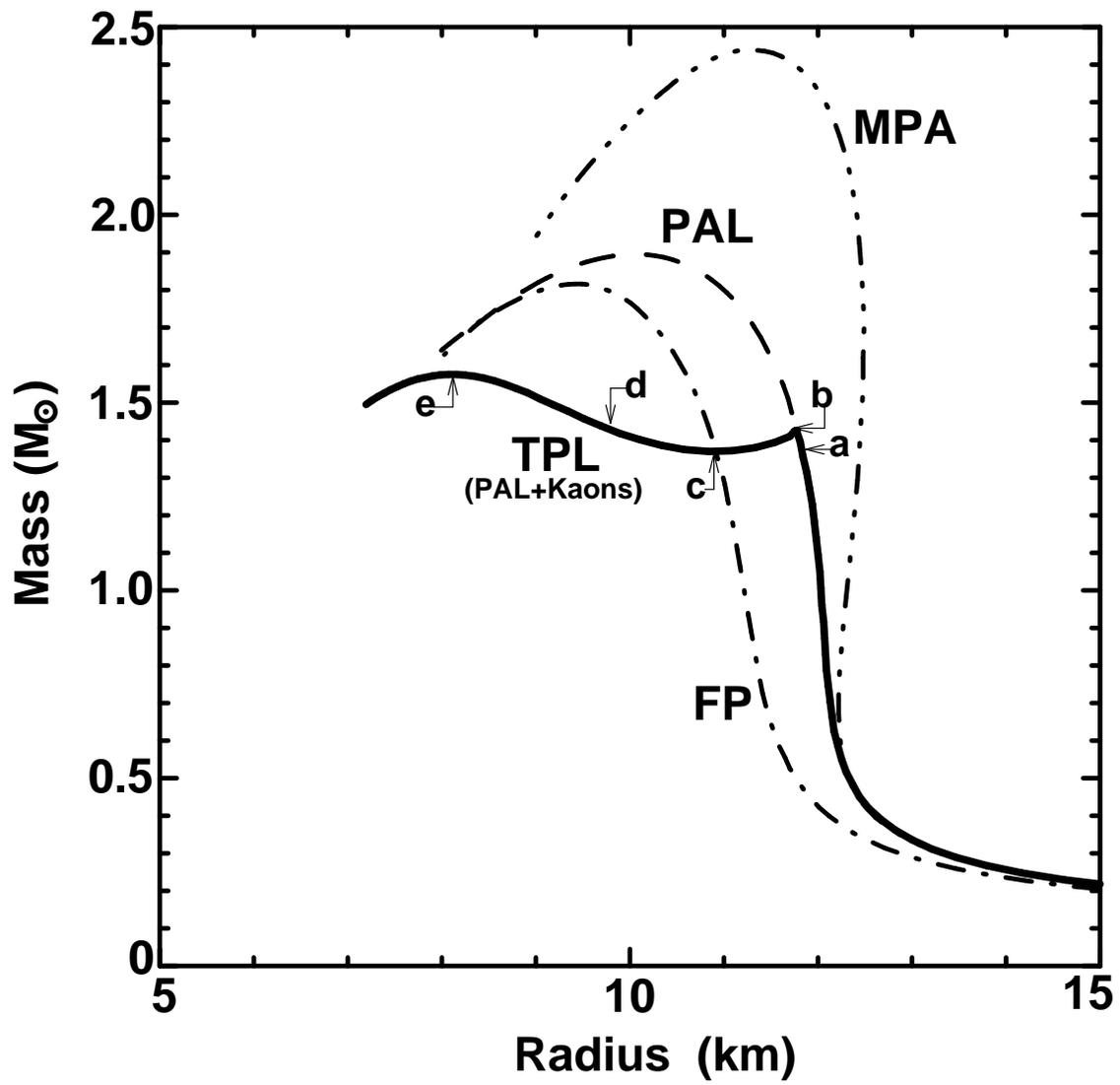
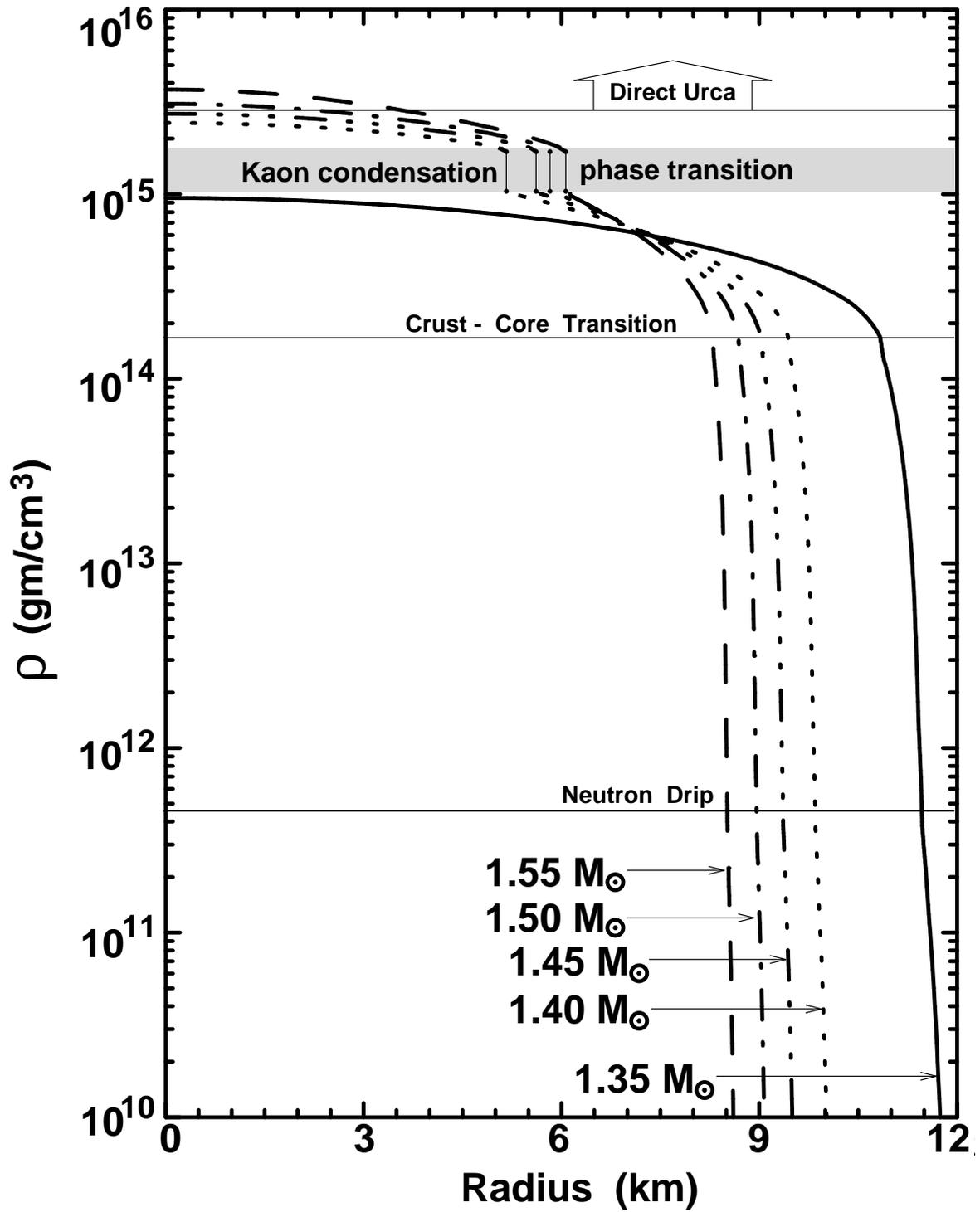
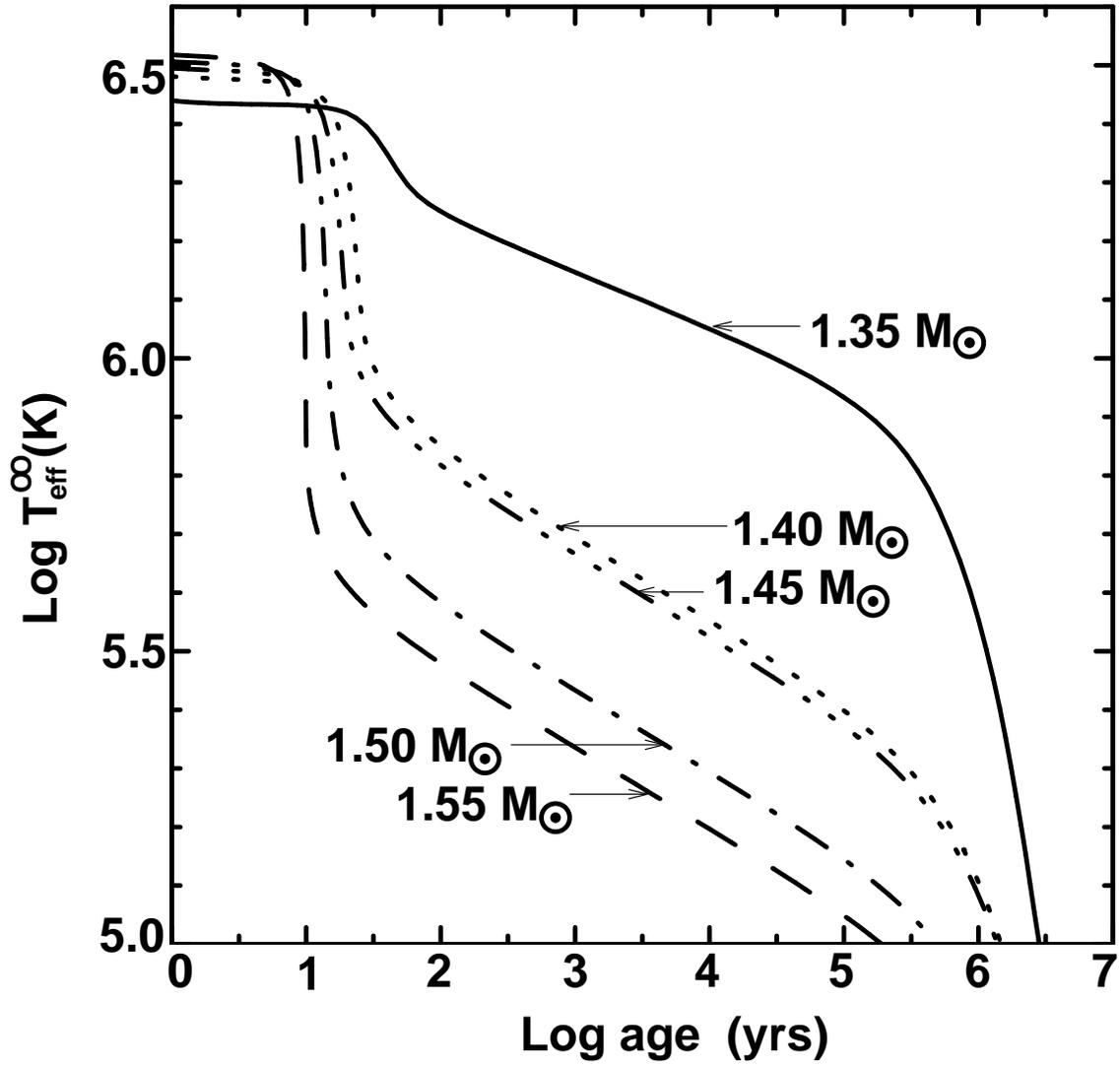
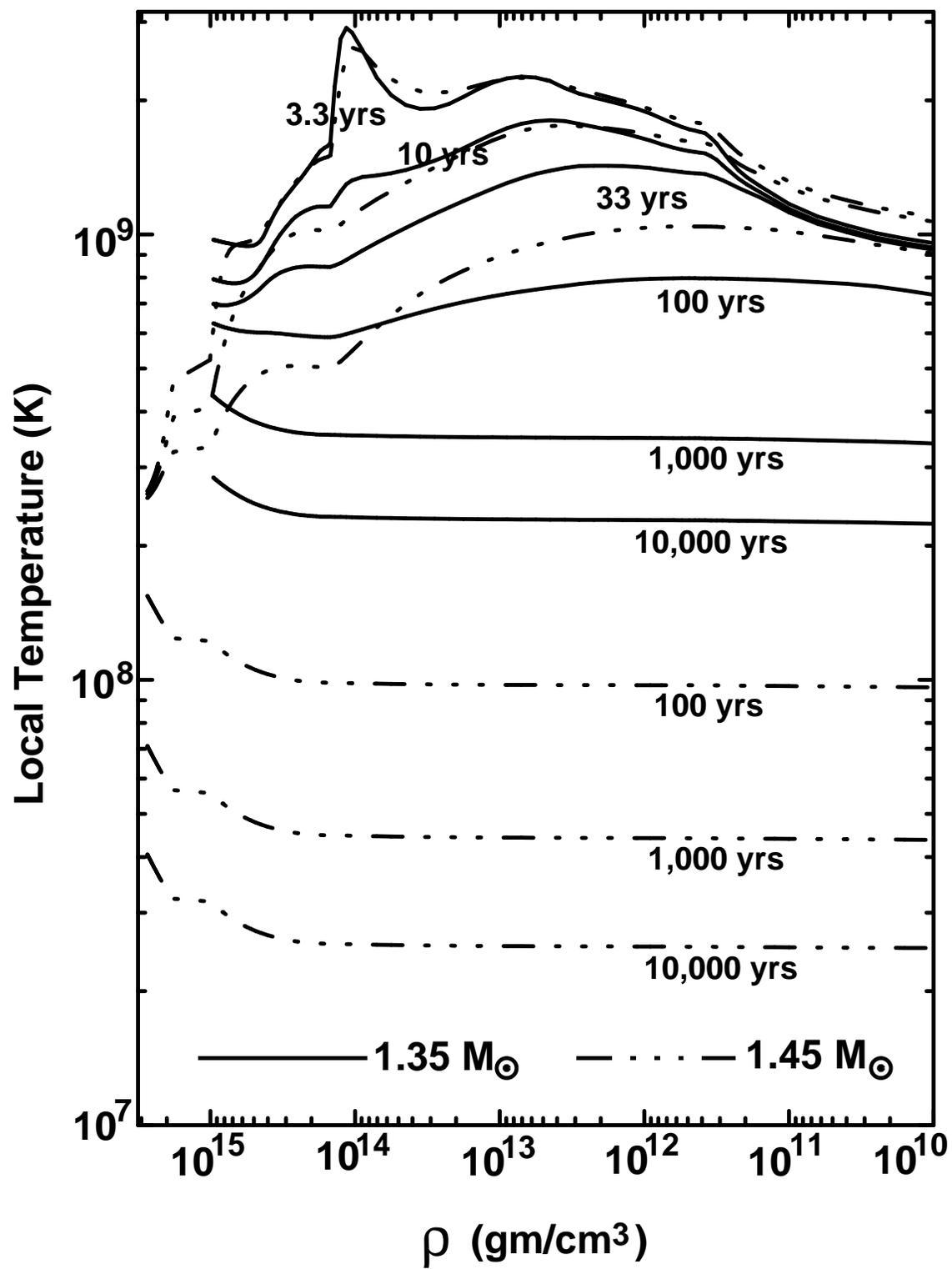
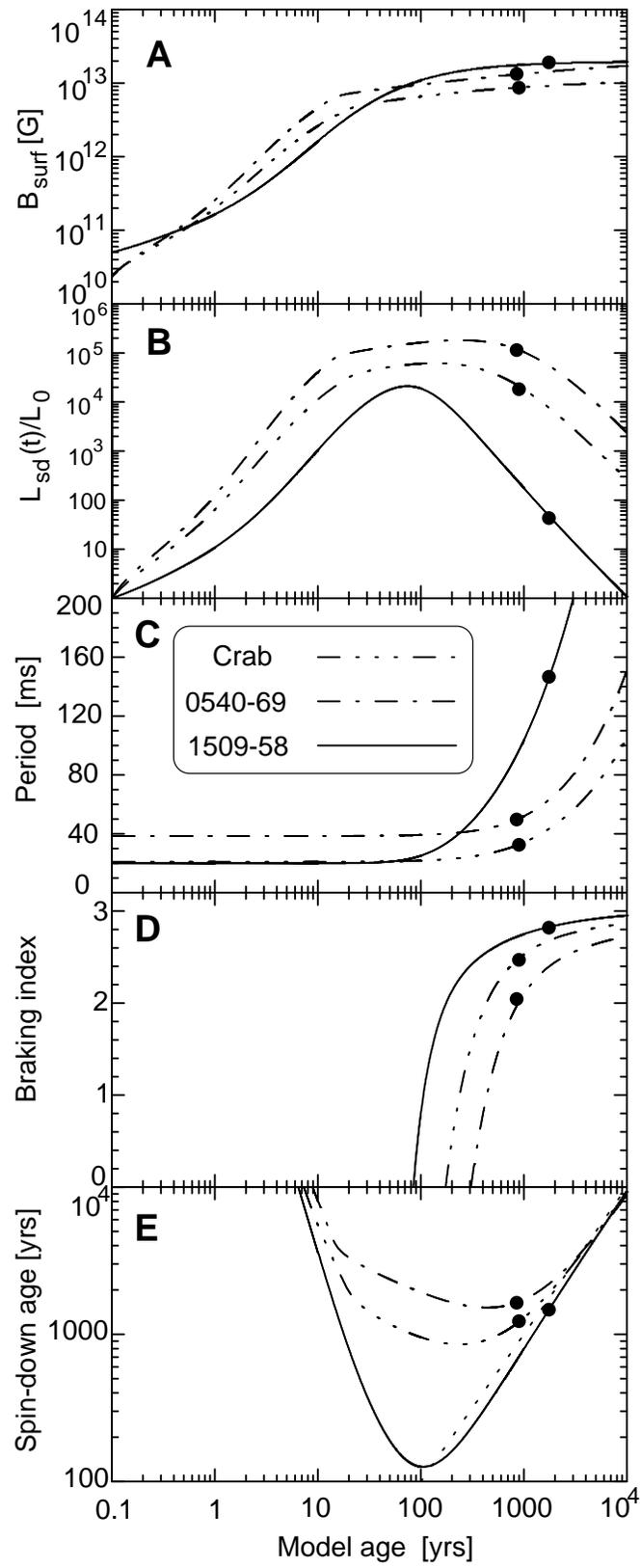
